\documentclass[aps, prl,twocolumn,superscriptaddress,showpacs,preprintnumbers,amsmath,amssymb]{revtex4}

\usepackage{graphicx}
\usepackage{dcolumn}
\usepackage{bm}

\begin{document}

\preprint{APS/123-QED}

\title{Generalized Spin Glass Relaxation}

\author{R M Pickup}
\affiliation{
School of Physics and Astronomy, University of Leeds, Leeds LS2 9JT, UK}
\author{R Cywinski}

\affiliation{
 School of Applied Sciences, University of Huddersfield, Huddersfield HD1 3DH, UK}

\email{r.cywinski@hud.ac.uk}

\author{C Pappas}
\affiliation{
Helmholtz Center Berlin for Materials and Energy, Glienickerstr, 100, 14109, Berlin, Germany}

\author{B Farago}
\affiliation{
Institut Laue Langevin, 6 rue Jules Horowitz, 38000 Grenoble, France} 

\author{P Fouquet}
\affiliation{
Institut Laue Langevin, 6 rue Jules Horowitz, 38000 Grenoble, France} 
 
\date{\today}

\begin{abstract}

Spin relaxation close to the glass temperature of \underline{Cu}Mn and \underline{Au}Fe spin glasses is shown, by neutron spin echo, to follow a generalized exponential function which explicitly introduces hierarchically constrained dynamics and macroscopic interactions. The interaction parameter is directly related to the normalized Tsallis non-extensive entropy parameter, $q$, and exhibits universal scaling with reduced temperature. At the glass temperature  $q= 5/3$ corresponding, within Tsallis' q-statistics, to a mathematically defined critical value for the onset of strong disorder and non-linear dynamics. \end{abstract}

\maketitle

The  dynamical correlations associated with the onset of a glassy state have been the focus of considerable interest and controversy over the last four decades, yet despite this attention such dynamics are still poorly understood. Above the glass temperature, $T_g$, the dynamics are generally characterized by a stretched exponential (Kohlrausch or KWW  \cite{kohl, phillips} ) relaxation, $\sim \exp [{-\left( t/\tau \right)^\beta ]} $. This form appears to be almost ubiquitous in nature, describing phenomena as diverse as human dynamics  \cite{Nakamura}, the conductivity close to the metal-insulator transition \cite{Jaroszyski}, the jamming transition \cite{Chaudhuri}, frustrated magnets \cite{Mutka}, experimental \cite{camp} and theoretical \cite{ogiel} spin glasses, as well as conventional glass forming liquids above the glass temperature \cite{Knaak, Richter_zorn}.

However the stretched exponential form does not adequately describe the relaxation of either structural or magnetic glasses  close to $T_g$ and below, where self-similarity in time (i.e. fractal behaviour in time) occurs. This is demonstrated in  Monte Carlo calculations by Ogielski \cite{ogiel}, based upon a 3d  $\pm$J Ising spin glass model, which show that close to the spin glass temperature the time dependent spin autocorrelation function should take a phenomenological modified Kohlrausch form which incorporates a power law dependence: 
\begin{equation}
\label{ogielski}
q \left( t\right)=\left<S_i \left(0\right).S_i \left( t \right) \right> \propto t^{-x}exp^{-\left(t/\tau \right)^\beta}
\end{equation}
in which $\beta $ increases from 1/3 at $T_g$ to 1 at $\sim 4T_g$, $x$ increases from 0 below $T_g$ to 0.5 at high temperatures and the relaxation time, $\tau$, diverges at $T_g$. This prediction is also largely supported by neutron spin echo (NSE) measurements \cite{papp}.

The origins and implications of such non-exponential relaxation remain the subject of some debate, not least because it can be demonstrated that Kohlrausch-like relaxation may arise from either a statistical distribution of independent (parallel) relaxation channels, or from more complex hierarchically constrained dynamics \cite{palmer}. For a detailed physical insight it is necessary to explore global models of glassy relaxation which attempt intrinsically to embody both distributed dynamics and the interactions that may lead to hierarchical relaxation. 

Such a model was introduced by Weron \cite{weron} in an attempt to explain the apparently universal power law for dielectric relaxation. Weron's rigorous probabilistic approach, based on the cluster model of Dissado and Hill \cite{dissado}, considers a hierarchical progression of relaxation which results in a continuously changing energy landscape. The hierarchy is introduced through the formation of finite clusters which arise from interactions between relaxing dipoles, with each cluster being  represented by an effective dipole related to its internal structure. The time taken for polarization fluctuations to reach equilibrium is a random variable that for each relaxing dipole depends upon two other random variables, namely the waiting time and the dissipation rate. Through these variables Weron accounts for the effects of both inter-cluster and intra-cluster interactions with the characteristic timescale of any relaxing entity being restricted by the structural reorganization of the surrounding clusters and derives a generalized  relaxation function
\begin{equation}
\label{Weron}
\varphi(t)=\left[ 1+k \left( t/\tau \right) ^\beta \right] ^{-1/k}
\end{equation}
where $\beta$ is associated with the fractal geometry of the system and $k (>0)$ is an effective interaction parameter related to the waiting time and providing a measure of the relative contribution of hierarchichal relaxation processes. Conveniently, the Weron power law reduces to the Kohlrausch form in the limit $k\rightarrow0$, in which case $0 < \beta \leq 1$  has precisely the same meaning as before, with the limit $\beta \rightarrow 1$ implying simple Debye (exponential) relaxation. 

Phenomenologically, the Weron model can be readily extended to the spin glass transition problem yet, despite its rigor and elegance, this approach has not previously been used in the analysis of spin relaxation. We have therefore investigated the applicability of this generalized relaxation function to spin glasses by analysing neutron spin echo  spectra collected from the archetypal metallic systems   Cu$_{1-x}$Mn$_{x}$ (x=0.1, 0.16, 0.35) and by revisiting our previously measured NSE spectra from Au$_{1-x}$Fe$_{x}$ (x=0.14)\cite{papp}.  We have also explored the interconnections between the probabilistic Weron model and a more general approach based on non-extensive thermodynamics introduced by Tsallis \cite{tsall} to describe highly disordered systems governed by L\'{e}vy-stable distributions. 

Cu$_{1-x}$Mn$_{x}$ samples with x=0.1, 0.16 and 0.35 were prepared by argon arc-melting and subsequent cold-rolling. The resulting disks were homogenised at 900$^\circ$C before being quenched directly into water. The samples were found to have glass temperatures, T$_g$, of 45K, 74K and 153K respectively. The Cu$_{1-x}$Mn$_{x}$ NSE spectra were collected using the IN11C spectrometer \cite{faragoIN11} at the Institut Laue Langevin (Grenoble) at $Q=4 nm^{-1}$ with an incoming wavelength of $0.55nm$. Our previously reported spectra  from Au$_{0.86}$Fe$_{0.14}$ (T$_g$=41K)  were obtained on IN15 at $Q=0.4 nm^{-1}$ and a  wavelength of $0.8nm$ \cite{papp}.

\begin{figure}[!htp]
\includegraphics[width=85mm]{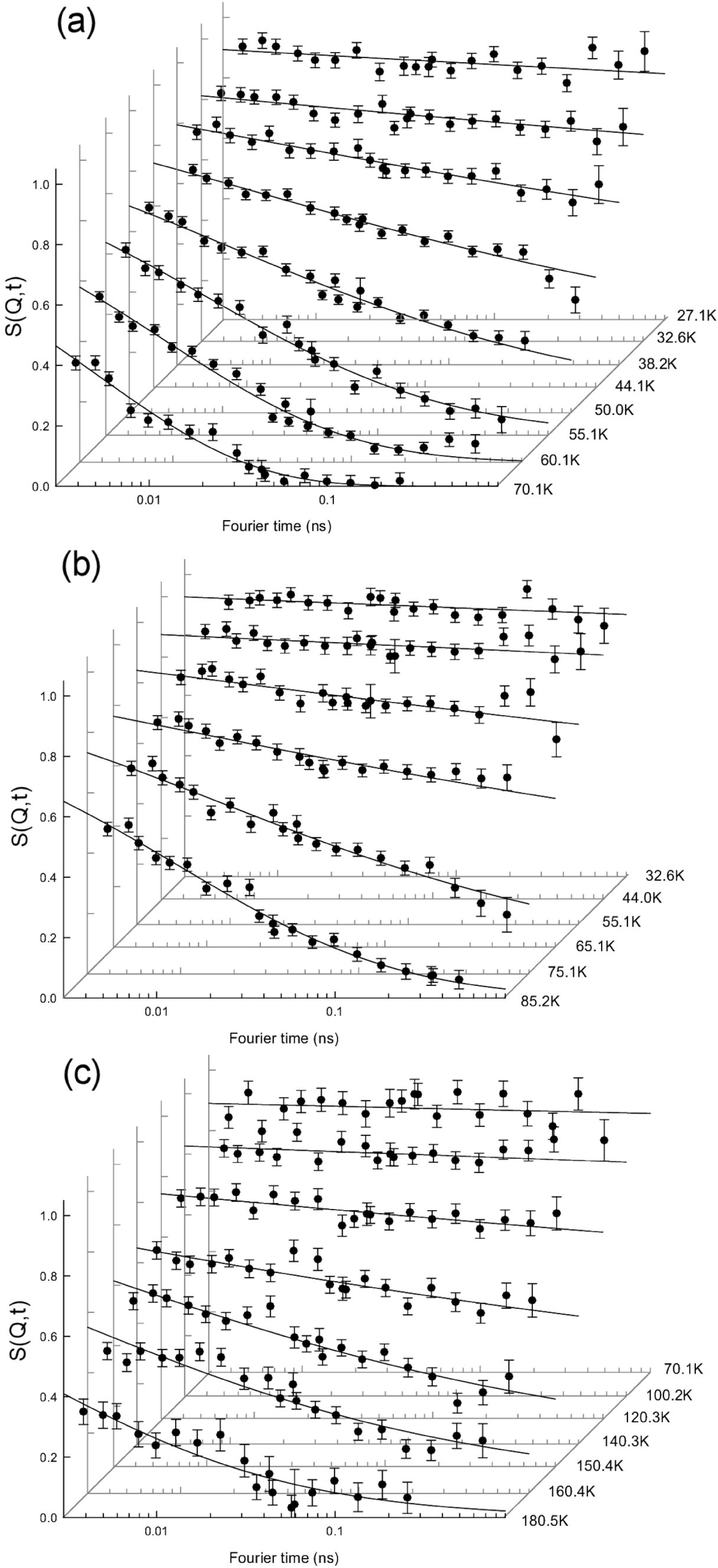} 
\caption{Temperature dependence of the NSE spectra of (a) Cu$_{0.9}$Mn$_{0.10}$ (b) Cu$_{0.84}$Mn$_{0.16}$ and (c) Cu$_{0.65}$Mn$_{0.35}$ at $Q=4nm^{-1}$. In each case the solid lines represent fits of the Weron function, eq(\ref{Weron}), to the data.}
\label{fig1}
\end{figure}

\begin{figure}[!htp]
\includegraphics[width=85mm]{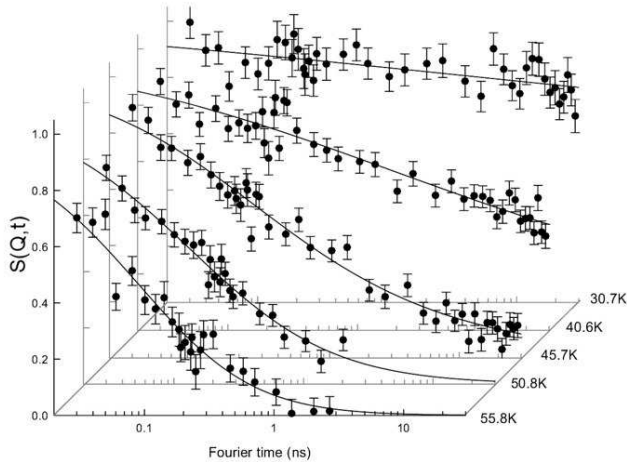}
\caption{Temperature dependence of the NSE spectra of Au$_{0.86}$Fe$_{0.14}$ at $Q=0.4nm^{-1}$. The solid lines are the associated fits of the Weron function, eq(\ref{Weron}).}
\label{fig2}
\end{figure}

\begin{figure}[!htp]
\includegraphics[width=85mm]{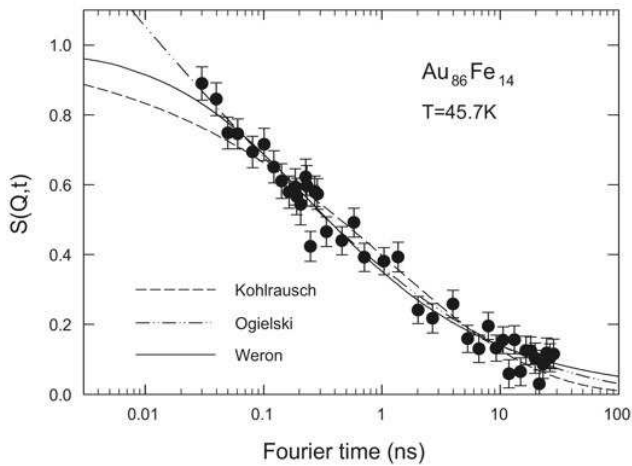}
\caption{The NSE spectra of Au$_{0.86}$Fe$_{0.14}$ at 45.7K with associated least squares fits of the Kohlrausch, Ogielski and Weron relaxation functions described in the text.}
\label{fig3}
\end{figure}

All NSE spectra were independent of scattering vector, $Q$: Characteristic results are shown in figures \ref{fig1} and \ref{fig2}. In  figure \ref{fig3} we show a direct comparison of least squares fits of the Kohlrausch stretched exponential, Ogielski (eq.\ref {ogielski}) and  Weron  (eq.\ref {Weron}) functions to the relaxation of  Au$_{0.86}$Fe$_{0.14}$  at 45.7 K. Whilst both of the latter functions afford equally acceptable descriptions over the Fourier time range covered by the data, the Ogielski function  must be modified to avoid an unphysical increase to values greater than unity at short times \cite{Keren}. In comparison, the Weron function is able to describe all spectra, from the high temperature simple exponential limit to the complex power law decay found below T$_g$, with consistent and physically meaningful parameters.

The spectra of Au$_{0.86}$Fe$_{0.14}$  give a temperature dependent $\beta$, decreasing from 1 at high temperatures to approximately 0.3 at the glass temperature, which  implies a continuing evolution of the geometric and dynamical fractal character of the spin clusters. The  \underline{Cu}Mn spectra cover a smaller dynamic range, and equally valid fits can be obtained with a temperature independent $\beta \sim 0.5 $ or with  $\beta$ varying with temperature as found for \underline{Au}Fe.  However, the most informative parameter  is the interaction parameter, $k$, which diverges as the spin glass temperature is approached. We suggest that $k$ provides an interesting and novel insight into the underlying thermodynamics which drive the non-exponential spin glass relaxation, and closely links such relaxation to that observed in other disordered systems.

A physical interpretation of the Weron function is provided by Tsallis' generalisation of Boltzman-Gibbs thermostatistics for complex and multifractal systems \cite{tsall}. In this context Tsallis introduces the concept of non-extensive entropy, proposing that the total entropy of any self-organising, strongly interacting systems may be either greater or less than the sum of the entropies of the individual components of the system. 

According to the Tsallis model, summing the entropy of two independent systems, $A$ and $B$, gives
\begin{equation}
\label{tsallis1}
\frac{S_{q}(A+B)}{k_B}=\frac{S_{q}(A)}{k_B}+\frac{S_{q}(B)}{k_B}+(1-q)\frac{S_{q}(A)}{k_B}.\frac{S_{q}(B)}{k_B}
\end{equation}
where $q$ is the so-called non-extensivity parameter. For $q>1$ the entropy is sub-extensive whilst for $q<1$ it is super-extensive. Standard Boltzmann-Gibbs entropy ($S=k_B\sum_i p_ilnp_i$) is recovered as $q\rightarrow1$. In general, the Tsallis total entropy may therefore be written as
\begin{equation}
\label{tsallis2}
S_{q}=k_B \frac{\displaystyle{1-\sum_i p^{q}_i}}{q-1}
\end{equation}

Brouers and Sotolongo-Costa \cite{soto} demonstrated a direct relation between Tsallis' non-extensive entropy and Weron's probabilistic relaxation model, taking as their starting point the  conventional weighted distribution of exponential relaxation processes
\begin{equation}
\label{distribution}
\varphi(t)=\int^{\infty}_{0}f(\tau)exp^{-t/\tau}d\tau
\end{equation}

Here the relaxation time, $\tau$, scales with the characteristic volume $v$ of the relaxing clusters according to  $\tau=v^{1/\beta}$ and  $0<\beta \leq 1$ is related to the fractal geometry of the system, as in the Weron analysis. By maximising the Tsallis non-extensive entropy they obtain a (normalised) cluster size distribution function
\begin{equation}
\label{distributionv}
f(v)=\left[1-v{\frac{1-q}{2-q}}\right]^{\frac{1}{1-q}}
\end{equation}
where $1\leq q < 2$. This distribution function belongs to the family of L\'{e}vy-stable distributions characterized by asymptotic power law tails \cite{Levy1, Levy2} and the non-extensivity parameter q quantifies  the statistics of the system. An analysis of eq. \ref{distributionv} shows that the moments are defined only for 1$<$q$<$3/2, whereas for q$\geq$3/2 all moments, including the mean cluster size, diverge. However, within the modified q-statistics  introduced by Tsallis finite moments are obtained only for  q$<$5/3  \cite{Prato}. The critical value q=5/3 marks the transition to the limit of strong disorder, where most statistical weight is in the wings of the distribution and the  macroscopic behavior is governed by a high number of highly improbable collective events. 

The relaxation function deduced from the  cluster size distribution function takes a form identical to that proposed by Weron, subject to the simple substitutions $k=(q-1)/(2-q)$ 
\begin{equation}
\label{generalised}
\varphi(t)= \left[1+\left(\frac{q-1}{2-q}\right).\left(\frac{t}{\tau}\right)^\beta \right]^{-\frac{2-q}{q-1}}
\end{equation}

The values of the interaction parameter, $k$, obtained from fits of the Weron function of eq. \ref{Weron} to the \underline{Au}Fe and \underline{Cu}Mn NSE spectra can thus be related directly to the sub-extensivity parameter, $q$. 
\begin{figure}[!htp]
\includegraphics[width=85mm]{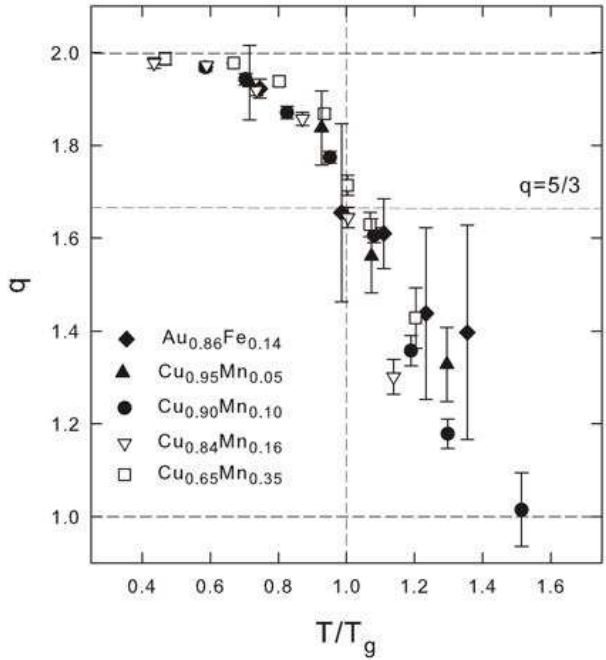}
\caption{The Tsallis sub-extensivity parameter, $q$, obtained from the fits of eq(\ref{Weron}) to NSE spectra of the \underline{Au}Fe and \underline{Cu}Mn samples, as function of reduced temperature. The values of $q$ for the Cu$_{0.95}$Mn$_{0.05}$ sample have been obtained by fitting the previously published data of ref \cite{mezei}.}
\label{fig4}
\end{figure}

As shown in  figure \ref{fig4}, the resulting $q$-values for the all the spin glass alloys discussed here, together with those obtained from a re-analysis of the earlier published NSE measurements \cite{mezei} on the more dilute Cu$_{0.95}$Mn$_{0.05}$, exhibit the same scaling with reduced temperature, ($T/T_g$). At first sight this is surprising as the fractal character of dilute and concentrated \underline{Cu}Mn  spin glasses and of \underline{Au}Fe is  expected to be quite different: Although an oscillatory RKKY  magnetic exchange dominates the interactions in both systems, in \underline{Au}Fe the Fe atoms are known to cluster, whilst in CuMn the Mn atoms anticluster. Correspondingly ferromagnetic percolation is found at 15at\%Fe in \underline{Au}Fe, but incipient antiferromagnetism is not evident in \underline{Cu}Mn until ~70at\%Mn \cite{Mydosh}. It is a remarkable property of phase transitions that critical phenomena do not depend on microscopic details but on the topology (i.e. interactions and dimensionality) of the systems. 

At high temperatures $q$ is close to 1, indicating an essentially independent, parallel, relaxation of cluster moments. q increases continuously with decreasing temperature and the first limit at $q$=3/2 is reached at $\sim$ 1.2 T$_g$. The most remarkable result, clearly demonstrated by figure \ref{fig4}, is  that the  spin glass transition temperature corresponds to the limit $q \sim$ 5/3.  If  T$_g$ marks the transition to self-similarity, as suggested by MC simulations \cite{ogiel} and experimental findings \cite{papp}, then the relaxation  at T$_g$ should asymptotically follow  the  power law   (t/$\tau$)$^{-y}$ with y=$\beta \times$(2-q)/(q-1). Using $\beta \sim$ 0.3 and q =5/3 we obtain an exponent y= 0.15, which  is very close to y$\sim$0.13, the value found previously for \underline{Au}Fe and CuMn 5\% \cite{papp}.

The correspondence between T$_g$ and a mathematically defined critical value for the extensivity parameter q, which defines a transition to complex non-linear dynamics, is new and has profound repercussions. It appears that spin glass systems evolve from a state characterised by  conventional, extensive Boltzmann-Gibbs statistics at high temperatures to one of extreme sub-extensivity at $T_g$.  As  q increases long-range correlations become more and more important and above q=3/2 the classical description breaks down, necessitating the introduction of the Tsallis q-statistics.
 The strong disorder limit is reached at q=5/3, where the complex dynamics are governed by the  power law tails of the  L\'{e}vy stable cluster size distribution function (eq. \ref{distributionv}).  Large-amplitude aperiodic fluctuations and nonlinearity dominate this limit, which can be identified with the (spin) glass transition and with the point, where self-similar relaxation sets-in.  Significantly, this evolution  appears to be universal, at least for the fundamentally different \underline{Au}Fe and \underline{Cu}Mn spin glasses discussed here suggesting that  spin glasses  are  very similar to many other complex disordered systems such as financial markets,  earthquakes, turbulence or jamming which are governed by  by self-similarity and the underlining  L\'{e}vy stable distributions.

\begin{acknowledgments}
We wish to acknowledge Karina Weron for extremely enlightening discussions. RMP acknowledges the financial support of an EPSRC doctoral training award.
\end{acknowledgments}

\bibliography{generalisedrelaxation}

\begin{thebibliography}{23}
\expandafter\ifx\csname natexlab\endcsname\relax\def\natexlab#1{#1}\fi
\expandafter\ifx\csname bibnamefont\endcsname\relax
  \def\bibnamefont#1{#1}\fi
\expandafter\ifx\csname bibfnamefont\endcsname\relax
  \def\bibfnamefont#1{#1}\fi
\expandafter\ifx\csname citenamefont\endcsname\relax
  \def\citenamefont#1{#1}\fi
\expandafter\ifx\csname url\endcsname\relax
  \def\url#1{\texttt{#1}}\fi
\expandafter\ifx\csname urlprefix\endcsname\relax\def\urlprefix{URL }\fi
\providecommand{\bibinfo}[2]{#2}
\providecommand{\eprint}[2][]{\url{#2}}

\bibitem[{\citenamefont{Kohlrausch}(1847)}]{kohl}
\bibinfo{author}{\bibfnamefont{R.}~\bibnamefont{Kohlrausch}},
  \bibinfo{journal}{Ann. Phys. Lpz.} \textbf{\bibinfo{volume}{12}},
  \bibinfo{pages}{393} (\bibinfo{year}{1847}).

\bibitem[{\citenamefont{Phillips}(1996)}]{phillips}
\bibinfo{author}{\bibfnamefont{J.~C.} \bibnamefont{Phillips}},
  \bibinfo{journal}{Rep. Prog. Phys.} \textbf{\bibinfo{volume}{59}},
  \bibinfo{pages}{1133} (\bibinfo{year}{1996}).

\bibitem[{\citenamefont{Nakamura et~al.}(2007)\citenamefont{Nakamura, Kiyono,
  Yoshiuchi, Nakahara, Struzik, and Yamamoto}}]{Nakamura}
\bibinfo{author}{\bibfnamefont{T.}~\bibnamefont{Nakamura}},
  \bibinfo{author}{\bibfnamefont{K.}~\bibnamefont{Kiyono}},
  \bibinfo{author}{\bibfnamefont{K.}~\bibnamefont{Yoshiuchi}},
  \bibinfo{author}{\bibfnamefont{R.}~\bibnamefont{Nakahara}},
  \bibinfo{author}{\bibfnamefont{Z.~R.} \bibnamefont{Struzik}},
  \bibnamefont{and} \bibinfo{author}{\bibfnamefont{Y.}~\bibnamefont{Yamamoto}},
  \bibinfo{journal}{Phys. Rev. Lett.} \textbf{\bibinfo{volume}{99}},
  \bibinfo{pages}{138103} (\bibinfo{year}{2007}).

\bibitem[{\citenamefont{Jaroszyski and Popovi}(2006)}]{Jaroszyski}
\bibinfo{author}{\bibfnamefont{J.}~\bibnamefont{Jaroszyski}} \bibnamefont{and}
  \bibinfo{author}{\bibfnamefont{D.}~\bibnamefont{Popovi}},
  \bibinfo{journal}{Phys. Rev. Lett} \textbf{\bibinfo{volume}{96}},
  \bibinfo{pages}{037403} (\bibinfo{year}{2006}).

\bibitem[{\citenamefont{Chaudhuri et~al.}(2007)\citenamefont{Chaudhuri,
  Berthier, and Kob}}]{Chaudhuri}
\bibinfo{author}{\bibfnamefont{P.}~\bibnamefont{Chaudhuri}},
  \bibinfo{author}{\bibfnamefont{L.}~\bibnamefont{Berthier}}, \bibnamefont{and}
  \bibinfo{author}{\bibfnamefont{W.}~\bibnamefont{Kob}},
  \bibinfo{journal}{Phys. Rev. Lett.} \textbf{\bibinfo{volume}{99}},
  \bibinfo{pages}{060604} (\bibinfo{year}{2007}).

\bibitem[{\citenamefont{Mutka et~al.}(2006)\citenamefont{Mutka, Ehlers, Payen,
  Bono, Stewart, Fouquet, Mendels, Mevellec, Blanchard, and Collin}}]{Mutka}
\bibinfo{author}{\bibfnamefont{H.}~\bibnamefont{Mutka}},
  \bibinfo{author}{\bibfnamefont{G.}~\bibnamefont{Ehlers}},
  \bibinfo{author}{\bibfnamefont{C.}~\bibnamefont{Payen}},
  \bibinfo{author}{\bibfnamefont{D.}~\bibnamefont{Bono}},
  \bibinfo{author}{\bibfnamefont{J.~R.} \bibnamefont{Stewart}},
  \bibinfo{author}{\bibfnamefont{P.}~\bibnamefont{Fouquet}},
  \bibinfo{author}{\bibfnamefont{P.}~\bibnamefont{Mendels}},
  \bibinfo{author}{\bibfnamefont{J.~Y.} \bibnamefont{Mevellec}},
  \bibinfo{author}{\bibfnamefont{N.}~\bibnamefont{Blanchard}},
  \bibnamefont{and} \bibinfo{author}{\bibfnamefont{G.}~\bibnamefont{Collin}},
  \bibinfo{journal}{Phys. Rev. Lett.} \textbf{\bibinfo{volume}{97}},
  \bibinfo{pages}{047203} (\bibinfo{year}{2006}).

\bibitem[{\citenamefont{Campbell et~al.}(1994)\citenamefont{Campbell, Schenck,
  Herlach, Gygax, Amato, Cywinski, and Kilcoyne}}]{camp}
\bibinfo{author}{\bibfnamefont{I.~A.} \bibnamefont{Campbell}},
  \bibinfo{author}{\bibfnamefont{A.}~\bibnamefont{Schenck}},
  \bibinfo{author}{\bibfnamefont{D.}~\bibnamefont{Herlach}},
  \bibinfo{author}{\bibfnamefont{F.~N.} \bibnamefont{Gygax}},
  \bibinfo{author}{\bibfnamefont{A.}~\bibnamefont{Amato}},
  \bibinfo{author}{\bibfnamefont{R.}~\bibnamefont{Cywinski}}, \bibnamefont{and}
  \bibinfo{author}{\bibfnamefont{S.~H.} \bibnamefont{Kilcoyne}},
  \bibinfo{journal}{Phys. Rev. Lett.} \textbf{\bibinfo{volume}{72}},
  \bibinfo{pages}{1291} (\bibinfo{year}{1994}).

\bibitem[{\citenamefont{Ogielski}(1985)}]{ogiel}
\bibinfo{author}{\bibfnamefont{A.}~\bibnamefont{Ogielski}},
  \bibinfo{journal}{Phys.\ Rev. B.} \textbf{\bibinfo{volume}{32}},
  \bibinfo{pages}{7384} (\bibinfo{year}{1985}).

\bibitem[{\citenamefont{Mezei et~al.}(1987)\citenamefont{Mezei, Knaak, and
  Farago}}]{Knaak}
\bibinfo{author}{\bibfnamefont{F.}~\bibnamefont{Mezei}},
  \bibinfo{author}{\bibfnamefont{W.}~\bibnamefont{Knaak}}, \bibnamefont{and}
  \bibinfo{author}{\bibfnamefont{B.}~\bibnamefont{Farago}},
  \bibinfo{journal}{Phys. Rev. Lett.} \textbf{\bibinfo{volume}{58}},
  \bibinfo{pages}{571} (\bibinfo{year}{1987}).

\bibitem[{\citenamefont{Richter et~al.}(1992)\citenamefont{Richter, Zorn,
  Farago, Frick, and Fetters}}]{Richter_zorn}
\bibinfo{author}{\bibfnamefont{D.}~\bibnamefont{Richter}},
  \bibinfo{author}{\bibfnamefont{R.}~\bibnamefont{Zorn}},
  \bibinfo{author}{\bibfnamefont{B.}~\bibnamefont{Farago}},
  \bibinfo{author}{\bibfnamefont{B.}~\bibnamefont{Frick}}, \bibnamefont{and}
  \bibinfo{author}{\bibfnamefont{L.~J.} \bibnamefont{Fetters}},
  \bibinfo{journal}{Phys. Rev. Lett.} \textbf{\bibinfo{volume}{68}},
  \bibinfo{pages}{71} (\bibinfo{year}{1992}).

\bibitem[{\citenamefont{Pappas et~al.}(2003)\citenamefont{Pappas, Mezei,
  Ehlers, Manuel, and Campbell}}]{papp}
\bibinfo{author}{\bibfnamefont{C.}~\bibnamefont{Pappas}},
  \bibinfo{author}{\bibfnamefont{F.}~\bibnamefont{Mezei}},
  \bibinfo{author}{\bibfnamefont{G.}~\bibnamefont{Ehlers}},
  \bibinfo{author}{\bibfnamefont{P.}~\bibnamefont{Manuel}}, \bibnamefont{and}
  \bibinfo{author}{\bibfnamefont{I.~A.} \bibnamefont{Campbell}},
  \bibinfo{journal}{Phys.\ Rev. B.} \textbf{\bibinfo{volume}{68}},
  \bibinfo{pages}{054431} (\bibinfo{year}{2003}).

\bibitem[{\citenamefont{Palmer et~al.}(1984)\citenamefont{Palmer, Stein,
  Abrahams, and Anderson}}]{palmer}
\bibinfo{author}{\bibfnamefont{R.~G.} \bibnamefont{Palmer}},
  \bibinfo{author}{\bibfnamefont{D.~L.} \bibnamefont{Stein}},
  \bibinfo{author}{\bibfnamefont{E.}~\bibnamefont{Abrahams}}, \bibnamefont{and}
  \bibinfo{author}{\bibfnamefont{P.~W.} \bibnamefont{Anderson}},
  \bibinfo{journal}{Phys.\ Rev. Lett.} \textbf{\bibinfo{volume}{53}},
  \bibinfo{pages}{958} (\bibinfo{year}{1984}).

\bibitem[{\citenamefont{Weron}(1991)}]{weron}
\bibinfo{author}{\bibfnamefont{K.}~\bibnamefont{Weron}}, \bibinfo{journal}{J.
  Phys. C: Solid State Phys.} \textbf{\bibinfo{volume}{3}},
  \bibinfo{pages}{9151} (\bibinfo{year}{1991}).

\bibitem[{\citenamefont{Dissado and Hill}(1989)}]{dissado}
\bibinfo{author}{\bibfnamefont{L.~A.} \bibnamefont{Dissado}} \bibnamefont{and}
  \bibinfo{author}{\bibfnamefont{R.~M.} \bibnamefont{Hill}},
  \bibinfo{journal}{J. Appl. Phys.} \textbf{\bibinfo{volume}{66}},
  \bibinfo{pages}{2511} (\bibinfo{year}{1989}).

\bibitem[{\citenamefont{Tsallis et~al.}(1995)\citenamefont{Tsallis, Levy,
  Souza, and Maynard}}]{tsall}
\bibinfo{author}{\bibfnamefont{C.}~\bibnamefont{Tsallis}},
  \bibinfo{author}{\bibfnamefont{S.~V.~F.} \bibnamefont{Levy}},
  \bibinfo{author}{\bibfnamefont{A.~M.~C.} \bibnamefont{Souza}},
  \bibnamefont{and} \bibinfo{author}{\bibfnamefont{R.}~\bibnamefont{Maynard}},
  \bibinfo{journal}{Phys.\ Rev. Lett.} \textbf{\bibinfo{volume}{75}},
  \bibinfo{pages}{3589} (\bibinfo{year}{1995}).

\bibitem[{\citenamefont{Farago}(1997)}]{faragoIN11}
\bibinfo{author}{\bibfnamefont{B.}~\bibnamefont{Farago}},
  \bibinfo{journal}{Physica B} \textbf{\bibinfo{volume}{241}},
  \bibinfo{pages}{113} (\bibinfo{year}{1997}).

\bibitem[{\citenamefont{Keren et~al.}(2004)\citenamefont{Keren, Gardner,
  Ehlers, Fukaya, Segal, and Uemura}}]{Keren}
\bibinfo{author}{\bibfnamefont{A.}~\bibnamefont{Keren}},
  \bibinfo{author}{\bibfnamefont{J.~S.} \bibnamefont{Gardner}},
  \bibinfo{author}{\bibfnamefont{G.}~\bibnamefont{Ehlers}},
  \bibinfo{author}{\bibfnamefont{A.}~\bibnamefont{Fukaya}},
  \bibinfo{author}{\bibfnamefont{E.}~\bibnamefont{Segal}}, \bibnamefont{and}
  \bibinfo{author}{\bibfnamefont{Y.~J.} \bibnamefont{Uemura}},
  \bibinfo{journal}{Phys. Rev. Lett.} \textbf{\bibinfo{volume}{92}},
  \bibinfo{pages}{107204} (\bibinfo{year}{2004}).

\bibitem[{\citenamefont{Brouers and Sotolongo-Costa}(2003)}]{soto}
\bibinfo{author}{\bibfnamefont{F.}~\bibnamefont{Brouers}} \bibnamefont{and}
  \bibinfo{author}{\bibfnamefont{O.}~\bibnamefont{Sotolongo-Costa}},
  \bibinfo{journal}{Europhysics Letters} \textbf{\bibinfo{volume}{62}},
  \bibinfo{pages}{808} (\bibinfo{year}{2003}).

\bibitem[{\citenamefont{Nolan}(2009)}]{Levy1}
\bibinfo{author}{\bibfnamefont{J.~P.} \bibnamefont{Nolan}},
  \emph{\bibinfo{title}{Stable Distributions - Models for Heavy Tailed Data}}
  (\bibinfo{publisher}{Birkh\"auser}, \bibinfo{year}{2009}), \bibinfo{note}{in
  progress, Chapter 1 online at academic2.american.edu/$\sim$jpnolan}.

\bibitem[{\citenamefont{Barndorff-Nielsen
  et~al.}(2001)\citenamefont{Barndorff-Nielsen, Sidney, and Resnick}}]{Levy2}
\bibinfo{author}{\bibfnamefont{O.~E.} \bibnamefont{Barndorff-Nielsen}},
  \bibinfo{author}{\bibfnamefont{T.~M.} \bibnamefont{Sidney}},
  \bibnamefont{and} \bibinfo{author}{\bibfnamefont{I.}~\bibnamefont{Resnick}},
  \emph{\bibinfo{title}{L\'{e}vy Processes - Theory and applications}}
  (\bibinfo{publisher}{Birkh\"auser}, \bibinfo{year}{2001}).

\bibitem[{\citenamefont{Prato and Tsallis}(1999)}]{Prato}
\bibinfo{author}{\bibfnamefont{D.}~\bibnamefont{Prato}} \bibnamefont{and}
  \bibinfo{author}{\bibfnamefont{C.}~\bibnamefont{Tsallis}},
  \bibinfo{journal}{Phys. Rev. E.} \textbf{\bibinfo{volume}{60}},
  \bibinfo{pages}{2398} (\bibinfo{year}{1999}).

\bibitem[{\citenamefont{Mezei and Murani}(1979)}]{mezei}
\bibinfo{author}{\bibfnamefont{F.}~\bibnamefont{Mezei}} \bibnamefont{and}
  \bibinfo{author}{\bibfnamefont{A.~P.} \bibnamefont{Murani}},
  \bibinfo{journal}{J. Magn. Magn. Mater.} \textbf{\bibinfo{volume}{14}},
  \bibinfo{pages}{211} (\bibinfo{year}{1979}).

\bibitem[{\citenamefont{Mydosh}(1993)}]{Mydosh}
\bibinfo{author}{\bibfnamefont{J.~A.} \bibnamefont{Mydosh}},
  \emph{\bibinfo{title}{Spin Glasses: An Experimental Introduction}}
  (\bibinfo{publisher}{Taylor \& Francis, London}, \bibinfo{year}{1993}).

\end{thebibliography}

\end{document}